\documentclass[letterpaper, 10 pt, conference]{ieeeconf}
\IEEEoverridecommandlockouts
\usepackage{amsmath,amssymb,amsfonts}
\usepackage{graphicx}
\usepackage{hyperref}
\usepackage{verbatim}   
\usepackage{ntheorem}
\usepackage{algorithm}
\usepackage{algpseudocode}

\usepackage{mathtools}
\usepackage{siunitx}
\usepackage{tabularx}
\usepackage{pgfplots}
\usepackage{pgfplotstable}
\usepackage{multirow}
\usepgfplotslibrary{fillbetween}
\usepackage{booktabs}
\pgfplotsset{compat = newest}
\usetikzlibrary{arrows,positioning,shapes,intersections,external,patterns,calc,fit,decorations,decorations.markings} 
\usepackage{cite}

\newtheorem{rem}{Remark}

\newtheorem{cor}{Corollary}
\newtheorem{prop}{Proposition}

\newcommand\tran{\mkern-2mu\raise1.25ex\hbox{$\scriptscriptstyle\top\hspace{0.5mm}$}\mkern-3.5mu}
\newcommand{\R}{\mathbb{R}}

\newcommand{\N}{\mathbb{N}}
\newcommand{\C}{\mathcal{C}}

\newcommand{\D}{\mathcal{D}}

\newcommand{\bm}[1]{{\boldsymbol{#1}}}

\DeclareMathOperator{\diag}{diag}
\DeclareMathOperator{\var}{var}

\DeclareMathOperator{\Prob}{P}
\DeclareMathOperator{\prob}{p}
\newcommand{\GP}{\mathcal{GP}}
\newcommand{\z}{\bm z}

\newcommand{\x}{\bm x}

\newcommand{\dx}{\dot{\bm x}}

\newcommand{\f}{\bm{f}}

\renewcommand{\u}{\bm{u}}
\newcommand{\y}{\bm{y}}

\usepackage[noabbrev]{cleveref} 
\crefname{rem}{Remark}{Remarks}
\crefname{exam}{Example}{Examples}
\crefname{assum}{Assumption}{Assumptions}
\crefname{prop}{Proposition}{Propositions}
\crefname{propy}{Property}{Properties}
\crefname{cor}{Corollary}{Corollaries}
\crefname{lem}{Lemma}{Lemmas}
\crefname{section}{Section}{Sections}
\crefname{thm}{Theorem}{Theorems}
\crefname{alg}{Algorithm}{Algorithms}
\crefname{defn}{Definition}{Definitions}
\crefname{figure}{Fig.}{Fig.}
\Crefname{figure}{Figure}{Figures}
\crefname{equation}{}{}

\title{\LARGE \bf Gaussian Process Port-Hamiltonian Systems:\\ Bayesian Learning with Physics Prior}
\author{Thomas Beckers$^1$, Jacob Seidman$^2$, Paris Perdikaris$^3$, George J. Pappas$^1$\hfill% <-this % stops a space
\thanks{This work was supported by the AFOSR under grant FA9550-19-1-0265 (Assured Autonomy in Contested Environments)}%
\thanks{$^{1}$ are with the Department of Electrical and Systems Engineering, University of Pennsylvania, Philadelphia, PA 19104, USA, tbeckers@seas.upenn.edu, pappasg@seas.upenn.edu}%
\thanks{$^{2}$ is with the graduate program in Applied Mathematics and Computational Science, University of Pennsylvania, Philadelphia, PA 19104, USA, seidj@sas.upenn.edu}%
\thanks{$^{3}$ is with the Department of Mechanical Engineering and Applied Mechanics, University of Pennsylvania, Philadelphia, PA 19104, USA, pgp@seas.upenn.edu}
}

\begin{document}

\maketitle
\thispagestyle{empty}
\pagestyle{empty}

\begin{abstract}
Data-driven approaches achieve remarkable results for the modeling of complex dynamics based on collected data. However, these models often neglect basic physical principles which determine the behavior of any real-world system. This omission is unfavorable in two ways: The models are not as data-efficient as they could be by incorporating physical prior knowledge, and the model itself might not be physically correct. We propose Gaussian Process Port-Hamiltonian systems (GP-PHS) as a physics-informed Bayesian learning approach with uncertainty quantification. The Bayesian nature of GP-PHS uses collected data to form a distribution over all possible Hamiltonians instead of a single point estimate. Due to the underlying physics model, a GP-PHS generates passive systems with respect to designated inputs and outputs. Further, the proposed approach preserves the compositional nature of Port-Hamiltonian systems.
\end{abstract}

%%%%%%%%%%%%%%%%%%%%%%%%%%%%%%%%%%%%%%%%%%%%%%%%%%
%%%%%%%%%%%%%%%%%%%%%%%%%%%%%%%%%%%%%%%%%%%%%%%%%%
\section{Introduction}
The modeling and identification of dynamical systems is a crucial task in a broad range of domains such as physics, engineering, applied mathematics and medicine~\cite{derler2011modeling,close2001modeling,andrian2004vitro}. System identification has a long history in control as most control strategies are derived based on a model of the plant. Whereas classical models of physical systems are typically based on first principles, there is a recent shift to more data-driven modeling of complex dynamics to capture more details with reduced engineering effort. However, this paradigm shift poses new questions regarding the efficiency, interpretability, and physical correctness of these models~\cite{hou2013model}. By physical correctness we mean that the learned model respects physical principles such as conservation of energy and passivity. Including these physical principles in a data-driven approach is beneficial in several ways: The models 1) are more meaningful as they respect the postulates of physics, 2) come with increased interpretability, and 3) can be more data-efficient as satisfaction of physical axioms results in a meaningful inductive bias of the model \cite{karniadakis2021physics}.
% removed the phrase "to some manifold"

A large class of physical systems can be described by Hamiltonian mechanics, see~\cite{vilasi2001hamiltonian}. The main idea is that the total energy in the system is described by the Hamiltonian, a smooth scalar function of the system's generalized coordinates. 
\begin{figure}[t]
\begin{center}
\vspace{0.2cm}
	\input{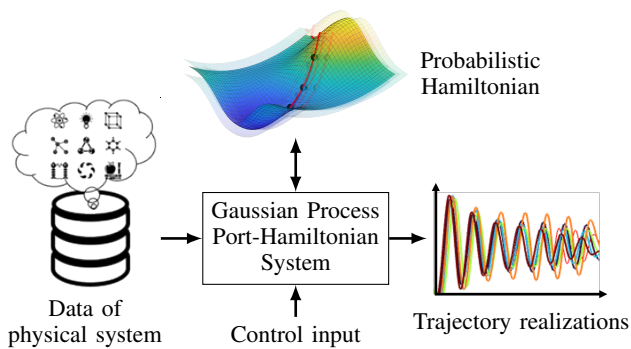}
	\vspace{-0.6cm}\caption{Gaussian Process Port-Hamiltonian systems allow to learn a probabilistic Hamiltonian based on collected data. The Port-Hamiltonian structure enables to include control inputs as well as dissipation. All sample trajectories of the GP-PHS model are physically correct in terms of the evolution of the total energy in the system.}\vspace{-0.7cm}
	\label{fig:gpphs_intro}
\end{center}
\end{figure}
 However, Hamiltonian systems are restricted to dynamics without external inputs and dissipative elements, which is rarely the case in real-world scenarios.

A composable framework to overcome these restrictions has been provided by Port-Hamiltonian systems (PHS), see~\cite{van2014port}. PHS leverage network approaches from electrical engineering and constitute a cornerstone of mathematical systems theory. While most of the analysis of physical systems has been performed within the Lagrangian and Hamiltonian framework, the network point of view is attractive for the modeling and simulation of complex physical systems with many connected components. PHS consider three kinds of ideal components: 1) energy-storing elements, 2) energy-dissipating (resistive) elements, and 3) energy-routing elements. It has been shown that PHS are able to model a range of multi-domain complex physical systems involving combinations of electrical, mechanical, electromechanical, chemical, hydrodynamical, thermodynamical, and quantum components, see \cite{van2000l2}. Therefore, PHS form an expressive class of systems that are able to model many real world physical scenarios of interest.

Results for the data-driven learning of Hamiltonian systems with neural networks have been established in~\cite{greydanus2019hamiltonian, bertalan2019learning}, and extended to PHS, see~\cite{desai2021port,nageshrao2015port}. However, standard neural network architectures can have difficulty generalizing from sparse data and have no uncertainty quantification of the outputs. In contrast, Bayesian models such as Gaussian processes (GP) can generalize well even for small training datasets and have a built-in notion of uncertainty quantification~\cite{rasmussen2006gaussian}. By using the Bayesian approach to learn a posterior \emph{distribution} of functions, we are able to both obtain point predictions of outputs as well as a quantified uncertainty. GPs have been recently used for learning generating functions for Hamiltonian systems~\cite{rath2021symplectic} or the dynamics of other structured systems~\cite{ridderbusch2021learning,raissi2018numerical,bhouri2021gaussian}, but they have not been applied to the class of Port-Hamiltonian systems.\\
\textbf{Contribution:} In this paper, we embed the powerful PHS representation into a Bayesian framework and propose Gaussian Process Port-Hamiltonian systems (GP-PHS) for learning physical systems based on state measurements. For this purpose, we introduce a PHS kernel for the GP that generates a distribution over PHS dynamics. The integration of physical principles allows the GP-PHS to be data-efficient and the resulting models are physically correct in the sense of the evolution of the system's energy. In contrast to existing PHS identifications methods, the probabilistic model includes all possible realizations of the PHS under the GP prior with respect to a finite set of data. We give a proof that the interconnection properties of PHS are preserved in GP-PHS which is a beneficial characteristic for system composition and control.

The remainder of the paper is structured as follows. We start with the introduction of PHS and GPs in~\cref{sec:def}, followed by proposing GP-PHS in~\cref{sec:mod}. Finally, a simulation highlights the benefits of GP-PHS in~\cref{sec:sim}.
%%%%%%%%%%%%%%%%%%%%%%%%%%%%%%%%%%%%%%%%%%%%%%%%%%
%%%%%%%%%%%%%%%%%%%%%%%%%%%%%%%%%%%%%%%%%%%%%%%%%%
\section{Problem Formulation}\label{sec:def}
In this section we briefly describe the class of Port-Hamiltonian systems, revisit GP regression, and introduce the problem setting. 
%%%%%%%%%%%%%%%%%%%%%%%%%%%%%%%%%%%%%%%%%%%%%%%%%%
\subsection{Port-Hamiltonian Systems}
Composing Hamiltonian systems with input/output ports leads to a  Port-Hamiltonian system (PHS) which is a dynamical system with ports that specify the interactions of its components. The dynamics of a PHS $\Sigma$ are described by\footnote{Vectors~$\bm a$ and vector-valued functions~$\bm f(\cdot)$ are denoted with bold characters. Matrices are described with capital letters. $I_n$ is the $n$-dimensional identity matrix and $0_n$ the zero matrix. The expression~$A_{:,i}$ denotes the i-th column of $A$. For a positive semidefinite matrix $\Lambda$, $\|x - y\|_{\Lambda}^2 = (x - y)^\top \Lambda (x-y)$.  $\R_{>0}$ denotes the set of positive real number whereas $\R_{\geq 0}$ is the set of non-negative real numbers. $\C^i$ denotes the class of $i$-th times differentiable functions. The operator $\nabla_\x$ with $\x\in\R^n$ denotes $[\frac{\partial}{\partial x_1},\ldots,\frac{\partial}{\partial x_n}]^\top$.}.
\begin{align}
\Sigma(J,R,H,G)\!\!=\!\!\begin{cases}
    \!\dx\!=\![J(\x)\!-\!R(\x)]\nabla_\x H( \x)\!+\!G(\x)\u\!\!\!\!\!\!\!\\
\!\y\!=\!G(\x)^\top \nabla_\x H(\x),\label{for:pch}
\end{cases}
\end{align}
with the state $\x(t)\in\R^n$ (also called energy variable) at time $t\in\R_{\geq 0}$, the total energy represented by a smooth function $H\colon\R^n\to\R$ called the  Hamiltonian, and the I/O ports $\u,\y\in\R^m$.

The matrix $J\colon\R^n\to\R^{n\times n}$ is skew-symmetric and specifies the interconnection structure and the matrix $R\colon\R^n\to\R^{n\times n},R=R^\top\succeq 0$ specifies the dissipation in the system. The contact with the environment is defined by the matrix $G\colon\R^n\to\R^{n\times m}$. The structure of the interconnection matrix~$J$ is typically derived from kinematic constraints in mechanical systems, Kirchhoff’s laws, power transformers, gyrators, etc.  Loosely speaking, the interconnection of the elements in the PHS is defined by $J$, whereas the Hamiltonian $H$ characterizes their dynamical behavior. The port variables~$\u$ and $\y$ are conjugate variables in the sense that their duality product defines the power flows exchanged with the environment of the system, for instance, currents and voltages in electrical circuits or forces and velocities in mechanical systems, see~\cite{van2000l2} for more information on PHS.
\begin{rem}
Port-Hamiltonian systems are a generalization of classical Hamiltonian systems but with the capability of including dissipation, input/output ports, and non-local coordinates. Thus, any Hamiltonian systems can be represented by a PHS~\cref{for:pch} with $J(\x)=\begin{bmatrix} 0 & I\\-I & 0\end{bmatrix},\,R=0,\,G=0$.
\end{rem}
%%%%%%%%%%%%%%%%%%%%%%%%%%%%%%%%%%%%%%%%%%%%%%%%%%
\subsection{Gaussian Process Regression}\label{sec:GPIntro}
Let~$(\Omega, \mathcal{F},P)$ be a probability triple with the probability space~$\Omega$, the corresponding~$\sigma$-algebra~$\mathcal{F}$ and the probability measure~$P$. Given a mean function $m_{\mathrm{GP}}: \R^n \rightarrow \R$ and a kernel function $k: \R^n \times \R^n \rightarrow \R$, a GP $\mathcal{GP}(m_{\mathrm{GP}}, k)$ is a stochastic process on a set $\mathcal{X} \subseteq \R^n$ with marginals defined as follows. Given a sample $f \sim \mathcal{GP}(m_{\mathrm{GP}}, k)$ and a finite collection of points $\x \subset \mathcal{X}$, the vector of pointwise evaluations $f(\x)$ is distributed according to the normal distribution $f(\x) \sim \mathcal N(m_{\mathrm{GP}}(\x), K(\x, \x))$, where $K(\x,\x)$ is a matrix of pairwise kernel evaluations at the points in $\x$. One of the main advantages of GPs is their combined use with Bayes' Theorem to provide tractable statistical inference for regression~\cite{rasmussen2006gaussian}. In this regard, consider the output $y\in\R$ of a continuous function~$f\colon \R^{n} \rightarrow \R$. The measurements might be affected by Gaussian noise such that $y = f(\x) + \eta$ with $\eta\sim\mathcal{N}(0,\sigma^2)$. We place a mean-zero GP prior with kernel $k$ on~$f$.  The training set $ \mathcal{D=} \left\lbrace X,Y \right \rbrace$ denotes the set of input data, $X=\left[\x^{\{1\}}, \x^{\{2\}}, \ldots, \x^{\{N\}}\right] \in \R^{n \times N}$ and measured output data, $Y=\left[y^{\{1\}}, y^{\{2\}}, \ldots, y^{\{N\}}\right]^\top \in \R^{N}$. For a test input $\x^{*}\in\R^n$, the estimation of $f(\x^{*})$ is provided by conditioning on the data which leads to the posterior 
\begin{align}
\mu\left(f\!\mid\!\x^{*}, \D\right)&\!=\! \bm{k}\left(\x^{*}, X\right)^{\!\top}\!K^{-1} Y,\label{for:gpmean}\\
    \var\left(f\mid\x^{*}, \D\right)&= k\left(\x^{*}, \x^{*}\right)-\bm{k}\left(\x^{*}, X\right)^{\top}K^{-1} \bm{k}\left(\x^{*}, X\right).\notag
\end{align}
The kernel $k$ is a measure for the correlation of two inputs~$(\x,\x^\prime)$. The function~$K\colon \R^{n\times N}\times  \R^{n\times N}\to\R^{N\times N}$ is called the Gram matrix whose elements are given by $K_{j',j}= k(X_{:, j'},X_{:, j})+\delta(j,j')\sigma^2$ for all $j',j\in\{1,\ldots,N\}$ with the delta function $\delta(j,j')=1$ for $j=j'$ and zero, otherwise. The vector-valued function~$\bm{k}\colon \R^n\times  \R^{n\times N}\to\R^N$, with the elements~$k_j = k(\x^*,X_{:, j})$ for all $j\in\{1,\ldots,N\}$, expresses the covariance between~$\x^*$ and the input training data $X$. The selection of the kernel and the determination of the corresponding hyperparameters can be seen as degrees of freedom of the regression. A popular kernel for GP models of dynamical systems is the squared exponential kernel, which we will use throughout the following sections. 
%%%%%%%%%%%%%%%%%%%%%%%%%%%%%%%%%%%%%%%%%%%%%%%%%%
\subsection{Problem Setting}\label{sec:ps}
We consider the problem of learning a physical system whose dynamics can be written in Port-Hamiltonian form~\cref{for:pch}. We assume that we have access to (potentially noisy) observations $\tilde{\x}(t)\in\R^n$ of the system state $\x(t)\in\R^n$ whose evolution over time $t\in\R_{\geq 0}$ follows 
\begin{align}
\label{for:pchobs}
        \dot \x&\!=\![J(\x)\!-\!R(\x)]\nabla_\x H( \x)\!+\!G(\x)\u
\end{align} 
starting at $\x(0)\in\R^n$. The Hamiltonian $H\in\C^\infty$ is assumed to be \emph{completely unknown} due to unstructured  uncertainties in the system typically imposed by nonlinear springs, physical coupling effects, or highly nonlinear electrical and magnetic fields. The parametric structures of the interconnection matrix $J\colon\R^n\to\R^{n\times n}$, dissipation matrix $R\colon\R^n\to\R^{n\times n}$ and I/O matrix $G\colon\R^n\to\R^{n\times m}$ are assumed to be known but the parameters themselves are assumed to be unknown. In detail, the unknown set of parameters will be described by $\bm{\varphi}_J\in\Phi_J\subseteq\R^{n_{\varphi_J}},{n_{\varphi_J}}\in\N$ for the estimated interconnection matrix $\hat{J}(x\vert\bm{\varphi}_J)\in\R^{n\times n} $, $\bm{\varphi}_R\in\Phi_R\subseteq\R^{n_{\varphi_R}},{n_{\varphi_R}}\in\N$ for the estimated dissipation matrix $\hat{R}(x\vert\bm{\varphi}_R)\in\R^{n\times n}$ and $\bm{\varphi}_G\in\Phi_G\subseteq\R^{n_{\varphi_G}},{n_{\varphi_G}}\in\N$ for the estimated I/O matrix $\hat{G}(x\vert\bm{\varphi}_G)\in\R^{n\times m}$.

\textbf{Problem Formulation:}  Given a dataset of time\-stamps $\{t_i\}_{i=1}^N$ and noisy state observations with inputs, i.e.~$\{\tilde \x(t_i),\bm{u}(t_i)\}_{i=1}^N$, learn a Hamiltonian $\hat{H}$ and parameters $\bm{\varphi_J}, \bm{\varphi_R}, \bm{\varphi_G}$ for the estimated interconnection, dissipation, and I/O matrix, $\hat J, \hat R, \hat G$, respectively, such that the observed data is described by the PHS
\begin{align}\label{for:pchmodel}
\begin{split}
        \dx&=[\hat{J}(\x)-\hat{R}(\x)]\nabla_\x\hat{H}( \x)+\hat{G}(\x)\u.\\
\end{split}
\end{align}
We assume that the data $\tilde \x(t_i)$ is generated according to $\tilde \x(t_i) = \x(t_i) + \bm{\eta}$ where $\x(t)$ comes from some true PHS~\cref{for:pchobs} with unknown $H$, unknown parameters of the matrices $J, R, G,$ and the noise $\bm{\eta}\in\R^n$ is distributed according to a zero-mean Gaussian  $\eta\sim\mathcal{N}(\bm{0},\diag[\sigma_1^2,\ldots,\sigma_n^2])$ where the variances $\sigma_1^2,\ldots,\sigma_n^2\in\R_{\geq 0}$ are unknown.
%%%%%%%%%%%%%%%%%%%%%%%%%%%%%%%%%%%%%%%%%%%%%%%%%%
%%%%%%%%%%%%%%%%%%%%%%%%%%%%%%%%%%%%%%%%%%%%%%%%%%
\section{Gaussian Process Port-Hamiltonian Systems}\label{sec:mod}
\begin{figure}[t]
\begin{center}
\vspace{0.2cm}
	\input{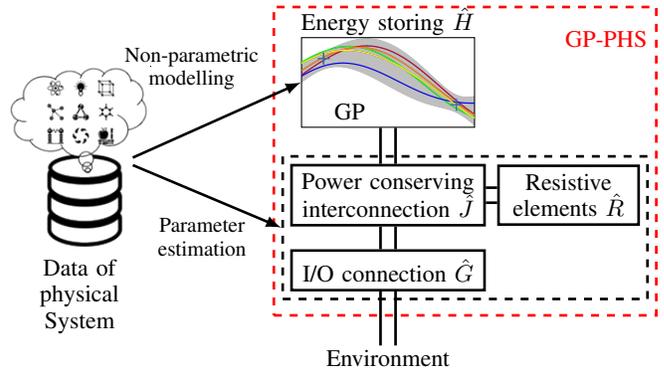}
	\vspace{-0.4cm}\caption{Block diagram of a Gaussian Process Port-Hamiltonian system.}\vspace{-0.6cm}
	\label{fig:gpphs}
\end{center}
\end{figure}

% In contrast to parametric models, data-driven models allow to let the data to speak for itself and, therefore, are highly interesting for the above stated problem where the functional form of the Hamiltonian is unknown. 
In this section, we propose Gaussian process Port-Hamiltonian systems (GP-PHS) whose structure is visualized in~\cref{fig:gpphs}. Starting with data of a physical system (on the left), we model the unknown Hamiltonian as nonparametric, probabilistic function. Additionally, unknown parameters of the interconnection matrix $J$, the dissipation matrix $R$ and the I/O matrix $G$ can be estimated exploiting the Bayesian nature of the GP.
In contrast to parametric models our approach is beneficial in several ways. First, we do not rely on prior knowledge about the parametric structure of the Hamiltonian. For instance, instead of assuming a linear spring model and fitting Hooke's spring constant based on data, the nonparametric structure of the GP allows to learn complex, nonlinear spring models.

Second, the Bayesian nature of the GP enables the model to represent \emph{all possible} PHS under the Bayesian prior based on a finite number of data points. This is not only interesting from a model identification perspective, but this uncertainty quantification can also be explicitly useful for model-based control approaches. However, using a GP to learn the Hamiltonian raises three challenges:
(i) The most straightforward application of GP regression, as presented in~\cref{sec:GPIntro}, would require us to have pointwise evaluations of $H(\x)$ but we do not have access to measurements of the Hamiltonian directly. (ii) The Hamiltonian only appears in the dynamics $\dx$ after being differentiated and we have no direct measurements of $\dx$. (iii) We need to find a kernel such that samples of the GP with this kernel always (or at least almost surely) result in physically correct PHS dynamics.

We addresses these challenges in the following section where we present the general structure of GP-PHS followed by the training and prediction process.
\subsection{Modeling}
First, we place a GP prior on the estimated Hamiltonian $\hat{H}\sim \GP(0,k(\x,\x^\prime))$, where $k$ is the squared exponential kernel. This choice of the kernel results in sampled Hamiltonians which are smooth and possesses a universality property which allows us to approximate any continuous function with it, see~\cite{rasmussen2006gaussian}. To overcome the fact that state measurements depend only on the partial derivatives of the Hamiltonian and not on the Hamiltonian itself, we leverage the property that GPs are closed under affine operations. Thus, the partial derivative can be included in the GP.  The system dynamics are given as an affine transformation of the Hamiltonian in the PH model~\cref{for:pchmodel}, so we can introduce a modified GP prior
\begin{align}
    \dx&\sim \GP(\hat{G}(\x)\u,k_{phs}(\x,\x^\prime))\notag
\end{align}
on the state evolution of~\cref{for:pchmodel}. Here, we introduce the new PHS kernel $k_{phs}\colon\R^n\times\R^n\to\R^{n\times n}$ given by
\begin{align}\label{for:phskernel}
    k_{phs}(\x,\x^\prime)&=\sigma_f^2\hat{J}_R(\x\mid \bm{\varphi}_J,\bm{\varphi}_R)\Pi(\x,\x^\prime)\hat{J}_R^\top(\x^\prime\mid \bm{\varphi}_J,\bm{\varphi}_R)\notag\\
    \Pi_{i,j}(\x,\x^\prime) &= \frac{\partial }{\partial z_i \partial z_j}\exp(-\|\z- \z^\prime\|_{\Lambda}^2)\Big\vert_{\z=\x,\z^\prime=\x^\prime}
\end{align}
where $\Pi\colon\R^n\times\R^n\to\R^{n \times n}$ describes the Hessian of the kernel and $\hat{J}_R(\x\mid \bm{\varphi}_J,\bm{\varphi}_R)=\hat{J}(\x\mid \bm{\varphi}_J)-\hat{R}(\x\mid \bm{\varphi}_R)$ is used for notational simplification. Note, that the PHS kernel~\cref{for:phskernel} is a matrix kernel as it maps to $\R^{n \times n}$, see~\cite{alvarez2012kernels}. The hyperparameters of $k_{phs}$ are the signal noise $\sigma_f\in\R_{>0}$, the lengthscales $\Lambda=\diag(l_1^2,\ldots,l_n^2)\in\R_{>0}^{n}$, and the parameter vectors $\bm{\varphi}_J,\bm{\varphi}_R,\bm{\varphi}_G$. Next, we prove that a GP with PHS kernel generates valid samples of a PHS with probability 1.
\begin{prop}\label{prop:1}
    Consider a vector field $\f\colon\mathcal{X}\times\R^m\to\R^n$ sampled from a GP equipped with the PHS kernel~\cref{for:phskernel}, i.e., $\bm{f}\sim \GP(\hat{G}(\x)\u,k_{phs}(\x,\x^\prime))$. For all realizations of $\bm{f}$ in the sample space $\Omega$, the dynamics
    \begin{align}
    \begin{split}\label{for:phs}
        \dx&=\f(\x,\bm{u},\omega),\quad \omega\in\Omega\\
        \y&=\hat G(\x)^\top \nabla_\x \hat H(\x,\omega),
        \end{split}
    \end{align}
    describe a PH system that is almost surely passive with respect to the supply rate $\u^\top \y$ on a compact set $\mathcal{X}$.
\end{prop}
\begin{proof}
    As we place a GP with squared exponential kernel $k(\x,\x^\prime)=\sigma_f\exp(-\|\x- \x^\prime\|_{\Lambda}^2)$ on the estimated Hamiltonian $\hat H$, all realizations $\hat H(\x,\omega)$ with $\omega\in\Omega$ are smooth functions in $\x$, see~\cite[Section 4.2]{rasmussen2006gaussian}. Using the fact that every smooth function $\hat H: \R^n \to \R$ defines a Port-Hamiltonian vector field under the affine transformation $\hat H \mapsto \hat J_R \nabla \hat H + \hat G u$, see~\cite{maschke1993port}, all realizations $\hat H(\x,\omega)$ define Port-Hamiltonian vector fields. 
    
    To show passivity, we must first show that there exist a $c_0\in\mathbb{R}$ such that for almost all $\omega \in \Omega$, $\hat H(\x,\omega)\geq c_0$, see~\cite{van2014port}. This guarantees that infinite energy cannot be extracted from the system. As the mean and covariance of a GP with squared exponential kernel are bounded, see~\cite{beckers:cdc2016}, the standard deviation metric $d(\x,\x^\prime)=\var(\hat H(\x)-\hat H(\x^\prime))^{1/2}$ for all $\x,\x^\prime\in\R^n$ is totally bounded, and we get $d(\x, \x^\prime)< c_1$ for some $c_1 \in \mathbb{R}_{> 0}$. This, together with the sample path continuity resulting from the squared exponential kernel allows us to appeal to the central limit theorem for stochastic processes, and conclude that $\hat H$ is almost surely bounded, i.e,~$\Prob(\sup_{\x\in\R^n}\vert \hat H \vert<\infty)=1$ on a compact set $\mathcal{X}$. Thus, we have shown that there exists a $c_0 \in \R$ such that almost surely $\hat H > c_0$.  Further, we can now consider the time-derivative of the sampled Hamiltonian for a $\omega\in\Omega$
\begin{align}
    \dot{\hat H}(\x,\omega)&=\nabla_\x^\top \hat{H}(\x,\omega) \hat{J}_R(\x)\nabla_\x \hat{H}(\x,\omega)\notag\\
    &+\nabla_\x^\top \hat{H}(\x,\omega) \hat{G}(\x)\u\label{for:tdham}\\
    &=-\nabla_\x^\top \hat{H}(\x,\omega) \hat{R}(\x)\nabla_\x \hat{H}(\x,\omega)+\u^\top \y\leq \u^\top\notag \y,
\end{align}
where $\u^\top \y$ is the supply rate. As the dissipation matrix $\hat R$ is positive semi-definite by definition, equation \cref{for:tdham} can be simplified to $\dot{H}(\x)\leq \u^\top \y$. Thus, the change in the system's total energy $\hat H$ is less the supply rate with the difference of the dissipation energy.
\end{proof}
As consequence of~\cref{prop:1}, the PHS kernel allows us to build physically correct models in terms of conversation or dissipation of energy.  To model more complex systems, we often wish to combine two or more PHS through an interconnection.  It is known that the class of PHS are closed under such interconnections, see~\cite{cervera2007interconnection}. We will show that GP-PHS share the same characteristic.
\begin{prop}\label{prop:inter}
    Consider two GP-PHS~\cref{for:phs} described by $\Sigma(\hat J_1,\hat R_1, \hat H_1,\hat G_1)$ with input dimension $m_1\in\N$ and $\Sigma(\hat J_2,\hat R_2, \hat H_2,\hat G_2)$ with input dimension $m_2\in\N$, respectively. Let $(\u_1^{\text{c}},\bm{y}_1^{\text{c}})\in\R^{m_{\text{c}}\times m_{\text{c}}}$ and $(\u_2^{\text{c}},\bm{y}_2^{\text{c}})\in\R^{m_{\text{c}}\times m_{\text{c}}}$ be corresponding input/output pairs of dimension $m_{\text{c}}$ with $m_{\text{c}}\leq \min\{m_1, m_2\}$ for the connection of the two GP-PHS. Then, the energy preserving interconnection $\u_1^{\text{c}}= -\bm{y}_2^{\text{c}}$ and $\u_2^{\text{c}}= \bm{y}_1^{\text{c}}$ yields a GP-PHS.
\end{prop}
\begin{proof}
First, we start with the definition of two GP-PHS. The first system is given by
    \begin{align}
    \begin{split}
            \dx&=\hat J_{R1}(\x)\nabla_\x \hat H_1( \x,\omega_1)+\hat G_1(\x)\u_1\\
\y_1&=\hat G_1(\x)^\top \nabla_\x \hat H_1(\x,\omega_1),
    \end{split}\notag
\end{align}
with Hamiltonian $\hat H_1 \sim\GP(0,k(\x,\x^\prime))$, state $\x\in\R^{n_1}$, sample $\omega_1\in\Omega$, $\hat J_{R1}(\x)=\hat{J}_1(\x)-\hat{R}_1(\x)$, and input/output $\bm{u}_1,\bm{y}_1\in\R^{m_1}$. We separate the I/O matrix $\hat G_1$ into $\hat G_1^{\text{c}}(\x)\in\R^{n_1 \times m_c}$ and $\hat G_1^{\text{ex}}(\x)\in\R^{n_1 \times m_1-m_c}$ such that $\hat G(\x)\bm{u}_1=\hat G_1^{\text{c}}\u_1^{\text{c}}+\hat G_1^{\text{ex}}\u_1^{\text{ex}}$ with external input $\u_1^{\text{ex}}\in\R^{m_1-m_c}$. The output for connection $\y_1^\text{c}$ is given by $\y_1^\text{c}=\hat G_1^{\text{c}}(\x)^\top \nabla_\x \hat H_1(\x,\omega_1)$.\\
Analogously, the second system is defined by
\begin{align}
\begin{split}
    \dot{\bm{\xi}}&=\hat J_{R2}(\bm{\xi})\nabla_\bm{\xi} \hat H_2(\bm{\xi},\omega_2)+\hat G_2(\bm{\xi})\u_2\\
\y_2&=\hat G_2(\bm{\xi})^\top \nabla_\bm{\xi} \hat H_2(\bm{\xi},\omega_2),
\end{split}\notag
    \end{align}
with Hamiltonian $\hat H_2\sim\GP(0,k(\bm{\xi},\bm{\xi}^\prime))$, state $\bm{\xi}\in\R^{n_2}$, sample $\omega_2\in\Omega$, $\hat J_{R2}(\bm{\xi})=\hat{J}_2(\bm{\xi})-\hat{R}_2(\bm{\xi})$, and input/output $\bm{u}_2,\bm{y}_2\in\R^{m_2}$. The I/O matrix $\hat G_2$ is separated into $\hat G_2^{\text{c}}(\bm{\xi})\in\R^{n_2 \times m_c}$ and $\hat G_2^{\text{ex}}(x)\in\R^{n_2 \times m_2-m_c}$ such that $\hat G(\bm{\xi})\bm{u}_2=\hat G_2^{\text{c}}\u_2^{\text{c}}+\hat G_2^{\text{ex}}\u_2^{\text{ex}}$ with $\u_2^{\text{ex}}\in\R^{m_2-m_c}$. The output for connection $\y_2^\text{c}$ is given by $\y_2^\text{c}=\hat G_2^{\text{c}}(\bm{\xi})^\top \nabla_\bm{\xi} \hat H_2(\bm{\xi},\omega_2)$.\\
For the interconnection $\u_1^{\text{c}}= -\bm{y}_2^{\text{c}}$ and $\u_2^{\text{c}}= \bm{y}_1^{\text{c}}$, we get
    \begin{align}
        \begin{bmatrix} \dx\\ \dot{\bm{\xi}}\end{bmatrix}&\!\!=\!\!\underbrace{\begin{bmatrix} \hat J_{R1}(\x) & \!\!\!\!\!-\hat G_1^{\text{c}}(\x)[G_2^{\text{c}}(\bm{\xi})]^\top \\ \hat G_2^{\text{c}}(\bm{\xi})[\hat G_1^{\text{c}}(\x)]^\top & \hat J_{R2}(\bm{\xi})\end{bmatrix}}_{J(\x,\bm{\xi})-R(\x,\bm{\xi})}\!\!\begin{bmatrix} \nabla_\x \\ \nabla_\bm{\xi}\end{bmatrix} \!\hat{H}(\x,\bm{\xi},\bm{\omega})\notag\\
        &+\underbrace{[\hat G_1^{\text{ex}}(\x),\hat G_2^{\text{ex}}(\bm{\xi})]}_{G(\x,\bm{\xi})}\begin{bmatrix}\u_1^{\text{ex}}\\\u_2^{\text{ex}}\end{bmatrix}\notag\\
        \bm{y}&=[\hat G_1^{\text{ex}}(\x),\hat G_2^{\text{ex}}(\bm{\xi})]^\top\begin{bmatrix} \nabla_\x \\ \nabla_\bm{\xi}\end{bmatrix} \!\hat{H}(\x,\bm{\xi},\bm{\omega})\notag
    \end{align}
with $\bm{\omega}=[\omega_1,\omega_2]^\top$, $J(\x,\bm{\xi}),R(\x,\bm{\xi})\in\R^{n\times n},n=n_1+n_2$, and $G(\x,\bm{\xi})\in\R^{n\times m},m=m_1+m_2-2m_c$, and output $\y\in\R^n$. If we define the joint Hamiltonian, $\hat H\colon\R^{n_1}\times\R^{n_2}\to\R$, as $\hat{H}(\x,\bm{\xi},\bm{\omega})=\hat H_1(\x,\omega_1)+\hat H_2(\bm{\xi},\omega_2)$, then we see that
\begin{align}
\hat{H}\sim\GP\left(\bm{0},\begin{bmatrix} k_1(\x,\x^\prime) & 0\\0 & k_2(\bm{\xi},\bm{\xi}^\prime), \end{bmatrix}\right),\notag
\end{align}
where $k_i(\bm{z},\bm{z}^\prime)=\sigma_{f,i}\exp(-\|\bm{z} - \bm{z}\|_{\Lambda_i}^2)$ denotes the squared exponential kernel for $i=\{1,2\}$ with hyperparameters $\sigma_{f,1},\sigma_{f,2}\in\R_{>0}$ and  $\Lambda_1\in\R_{>0}^{n_1\times n_1},\Lambda_2\in\R_{>0}^{n_2\times n_2}$.
\end{proof}
\Cref{prop:inter} shows that the negative feedback interconnection of two GP-PHS lead to a GP-PHS again. This is in particular interesting for passivity-based control approaches as shown in, e.g.,~\cite{ortega2004interconnection}. Next, we describe the learning and sampling procedure.
%%%%%%%%%%%%%%%%%%%%%%%%%%%%%%%%%%%%%%%%%%%%%%%%%%%%%%%%%%
\subsection{GP-PHS Training}
As indicated in the problem formulation, we consider an observed system trajectory
\begin{align} \label{for:dataset0}
    \mathcal{D}=\{(t_1,\tilde{\x}(t_1), \bm{u}(t_1)),\ldots,(t_{N},\tilde{\x}(t_{N}), \bm{u}(t_N))\}
\end{align}
 of the unknown dynamics~\cref{for:pchobs} corresponding to measured inputs $\{\bm{u}(t_1), \ldots, \bm{u}(t_N)\}$, where $N$ is the total number of training pairs consisting of a time $t_i$ and a noisy state measurement $\tilde{\x}(t_i)$. The first challenge we address is how to extract data of the form $\{\x,\dx\}$ from \cref{for:dataset0} so we may apply GP regression with the PHS kernel to learn the dynamics \eqref{for:pchmodel}. To obtain the derivative $\dx$, we exploit again that GPs are closed under affine operations~\cite{adler2010geometry}. With a differentiable kernel $k$, we learn $n$ separated GPs on the training sets $\mathcal{D}_j=\{(t_i,\tilde{x}_j(t_i))\}_{i=1,\ldots,N}$ with $j=1,\ldots,n$. Thus, one GP for each dimension $j$ of the state $\x\in\R^n$ is trained. As in~\cref{sec:GPIntro}, we define an input and output matrix by
 \begin{align}
 \begin{split}\label{for:dataset}
     T&=[t_1,\ldots,t_N]\in\R^{1\times N}\\
     \tilde{X}&=[\tilde{\x}(t_1),\ldots,\tilde{\x}(t_N)]^\top\in\R^{N\times n}.
      \end{split}
 \end{align}
Using a differentiable kernel $k$, we obtain the distribution for each element of the \emph{state derivative} $\dot{x}_j\in\R$ by
 \begin{align}
     \mu(\dot{x}_j \mid t,\D)&=\bm{k}^{(1)}\left(t, T\right)^{\!\top}\!K^{-1} \tilde{X}_{:,j}\label{for:meanvarxdx}\\
     \var(\dot{x}_j \mid t,\D)&=\bm{k}^{(1,2)}(t, t)-\bm{k}^{(1)}\left(t, T\right)^{\top}K^{-1} \bm{k}^{(1)}\left(t, T\right),\notag
 \end{align}
where $K_{i,i'}= k(t_{i},t_{i'})+\delta(i,i')\sigma_j^2$ for all $i,i'\in\{1,\ldots,N\}$ with the delta function $\delta(i,i')=1$ for $i=i'$ and zero, otherwise. The term $k^{(l)}$ denotes the derivative of the kernel function $k$ with respect to the $l$-th argument, i.e., $k^{(1)}(t,T)_i= \frac{\partial}{\partial z}k(z, T_i)\big\vert_{z=t}$, $k^{(2)}(t,T)_i= \frac{\partial}{\partial z}k(t,z)\big\vert_{z=T_i}$, and ${k}^{(1,2)}(t,t)= \frac{\partial^2}{\partial z\partial z^\prime}k(z,z^\prime)\big\vert_{z=t,z^\prime=t}$ for $i=1,\ldots,N$. Further, an estimate of the state $\x(t_i)$ based on the noise state measurement $\tilde{\x}(t_i)$ is obtained by standard GP regression $\mu(x_j \mid t,\D_j)=\bm{k}\left(t, T\right)^{\top}K^{-1} \tilde{X}_{:,j}$. Thus, we can create a new dataset $\mathcal{E}=\{(\mu(\x \mid t_i,\D), \mu(\dx \mid t_i,\D)\}_{i=1,\ldots,N}$ that is suitable for the learning of an estimated Hamiltonian $\hat H$. 
\begin{rem}
For the sake of simplicity, we focus here on a single trajectory. However, the above procedure can be repeated for multiple trajectories and the dataset $\mathcal{E}$ can be updated accordingly. In fact, there exist efficient procedures for the updating of GPs with new data, e.g., see~\cite{huber2014recursive}.
\end{rem}
Further, the dataset is exploited for the estimation of the unknown parameters $\bm{\varphi}_J\in\Phi_J,\bm{\varphi}_R\in\Phi_R,\bm{\varphi}_G\in\Phi_G$ for the matrices $\hat{J}(x\vert\bm{\varphi}_J),\hat{R}(x\vert\bm{\varphi}_R)$, and $\hat{G}(x\vert\bm{\varphi}_G)$ of the PH model~\cref{for:pchmodel}. First, we write the dataset $\mathcal{E}$ as input and output data analogous to~\cref{for:dataset} by
\begin{align}
\begin{split}\label{for:data2}
     X&=[\mu(\x \mid t_1,\D),\ldots,\mu(\x \mid t_{N},\D)]\in\R^{n\times N}\\
     \dot{X}&=[\mu(\dx \mid t_1,\D),\ldots,\mu(\dx \mid t_{N},\D)]^\top\in\R^{N\times n}.
     \end{split}
\end{align}
As the unknown parameters are treated as hyperparameters of the GP, the marginal likelihood is given by $\prob(\dx\vert \bm{\varphi},X)=\int \prob(\dx\vert \bm{f},X)\prob(\bm{f}\vert \bm{\varphi},X)d\bm{f}$, where $\bm{\varphi}=[\bm{\varphi}_J^\top,\bm{\varphi}_R^\top,\bm{\varphi}_G^\top,\sigma_f,l_1,\ldots,l_n]^\top\in\R^{n_{\varphi_J}+n_{\varphi_R}+n_{\varphi_G}+n+1}$ contains the unknown parameters of the PHS and the kernel parameters $\sigma_f,l_1,\ldots,l_n\in\R_+$. With the Gaussian prior $\dot{X}\vert \bm{\varphi},X\sim\mathcal{N}\left(\bm{0},K_{phs}\right)$ we can compute the negative log marginal likelihood (NLML) of the data.

The matrix $K_{phs}\in\R^{n N\times n N}$ describes the covariance
\begin{align}\label{for:kphs}
    K_{phs}&=\begin{bmatrix}k_{phs}(X_{:,1},X_{:,1}) & \ldots & k_{phs}(X_{:,1},X_{:,N})\\\vdots & \ddots & \vdots\\ k_{phs}(X_{:,N},X_{:,1}) & \ldots & k_{phs}(X_{:,N},X_{:,N})\end{bmatrix}\notag\\
    &+\begin{bmatrix}I_n\var(\dx \mid t_1,\D) & 0 & 0\\ 0 & \ddots & 0\\ 0 & 0 & I_n\var(\dx \mid t_{N},\D)\end{bmatrix}
\end{align}
based on the matrix kernel $k_{phs}$ as shown in \cref{for:phskernel}. We use the posterior variance~\cref{for:meanvarxdx} of the estimated state derivative data~$\dot X$ as noise in the covariance matrix \cref{for:kphs}. This allows us to consider the uncertainty of the estimation in the modelling of the PHS. Then, we can compute the NLML 
\begin{align}
    -\log \prob(\dot{X}\vert \varphi,X)&\sim\dot{X}_0^\top K_{phs}^{-1} \dot{X}_0+\log\vert K_{phs} \vert,\label{for:loglik}
\end{align}
with the mean-adjusted output data  $\dot{X}_0=[[\mu(\dx \mid t_1,\D)-\hat{G}\bm{u}(t_1)]^\top,\ldots,[\mu(\dx \mid t_{N},\D)-\hat{G}\bm{u}(t_{N})]^\top]^\top$. Finally, the unknown (hyper)parameters $\bm\varphi$ can be computed by minimization of the NLML~\cref{for:loglik} via, e.g., a gradient-based method as the gradient is analytically tractable.
\subsection{Prediction}
Once the GP model is trained, we can draw samples from the posterior distribution using the joint distribution with mean-adjusted output data at a test states $\x^*\in\R^n$
\begin{align}
    \begin{bmatrix}\dot{X}_0\\\bm{f}(\x^*)\end{bmatrix}\!=\!\mathcal{N}\left(\bm{0},\begin{bmatrix}K_{phs} & k_{phs}(X,\x^*)\\k_{phs}(X,\x^*)^\top & k_{phs}(\x^*,\x^*)\end{bmatrix}\right),\notag
\end{align}
to obtain a vector field $\bm{f}$ for the GP-PHS model at these states.  However, for numerical integration purposes we will need to be able to access this vector field at an arbitrary number of points at arbitrary locations. An appealing idea is to sample the vector field $\dx$ along a grid of points and interpolate between them to create a callable vector field function for an ode-solver.  Unfortunately, we have no guarantee that this interpolation will produce a dynamics function that has a PH structure and respects the energy conservation/dissipation of the system. 

To overcome this issue, we propose to sample the estimated Hamiltonian $\hat H$ itself and create an interpolation of $\hat H$ instead of the vector field $\bm{f}$. In this case, we can use the fact that $\dot X_0$ and $\hat H$ are related through the linear transformation~$\hat H \mapsto \hat J_R \nabla \hat H$ and form the joint prior
\begin{align}\label{for:corH}
    \begin{bmatrix}\dot{X}_0\\\hat{H}(\x^*)\end{bmatrix}&\!=\!\mathcal{N}\left(\!\bm{0},\!\begin{bmatrix}K_{phs} & \bm{k}_{\dot{x}H}(X,\x^*)\\\bm{k}_{\dot{x}H}(X,\x^*)^\top & \!k_{HH}(\x^*,\x^*)\end{bmatrix}\right),
\end{align}
where the vector functions $\bm{k}_{\dot{x}H}(X,\x^*)\in\R^{nN\times 1}$ and $\bm{k}_{HH}(\x^*,\x^*)\in\R_+$ are constructed as shown in~\cref{sec:GPIntro} by the kernels
$k_{\dot{x}H}(\x,\x^\prime)=\hat{J}_R(\x)\nabla_x k_{HH}(\x,\x^\prime)$ and
$k_{HH}(\x,\x^\prime)=\sigma_f^2\exp(-\|\x - \x^\prime\|_{\Lambda}^2)$. We may then sample from the posterior of this distribution after conditioning on values of $\dot X_0$ to obtain values for $\hat H(\x^*)$. After interpolating~$\hat H$ along these points, the error of the interpolation will not affect the PHS properties as shown in the following.
\begin{cor}\label{cor:1}
Consider a GP-PHS trained on the dataset~\cref{for:dataset0} and a sampled Hamiltonian $\hat H(\cdot,\omega)$ by~\cref{for:corH} over a finite set $X^*\subset\R^n$ with $\omega\in\Omega$. Let $\hat{H}^*\colon\R^n\to\R$ be a smooth and bounded function approximator of $\hat H(\cdot,\omega)$. Then,~$
\dx=[\hat{J}(\x)-\hat{R}(\x)]\nabla_x \hat{H}^*(\x)+\hat{G}(\x)\u,\,\y=\hat G(\x)^\top \nabla_\x \hat{H}^*(\x)$ describes a Port-Hamiltonian system that is passive with respect to the supply rate $\u^\top \y$.
\end{cor}
\begin{proof}
    The proof results from~\cref{prop:1}.
\end{proof}
As consequence, we can use any function approximator that generates smooth and bounded functions to interpolate between the state-discrete samples of $\hat H(\x^*,\omega)$ over all $\x^*\in X^*$. Possible approaches are spline interpolation~\cite{de1978practical} or direct approximation of the GP posterior~\cite{wilson2020efficiently,beckers2020prediction}, among others. In~\Cref{alg:cap}, we summarize the steps to achieve samples of a GP-PHS.\vspace{-0.2cm}
\begin{algorithm}
\caption{Learning of GP-PHS}\label{alg:cap}
\begin{algorithmic}
\Require Trajectory $\mathcal{D}\leftarrow \{(t_i,\tilde{\x}(t_i))\}_{i=1,\ldots,N}$
\Require Control inputs $\{\bm{u}(t_i)\}_{i=1,\ldots,N}$
\State \textbf{Obtain $(\x,\dx)$ pairs}
\State Train $n$ independent GPs with $\mathcal{D}_j$~\cref{for:dataset}
\State Create $\mathcal{E}=\{(\mu(\x \mid t_i,\D), \mu(\dx \mid t_i,\D)\}_{i=1,\ldots,N}$
\State \textbf{Obtain GP-PHS model:}
\State Train GP-PHS with $\mathcal{E}$ and $\{\bm{u}(t_i)\}_{i=1,\ldots,N}$
\State Compute posterior variance $\var(\dx \mid T,\mathcal{D})$
\State Minimize NLML~\cref{for:loglik} to estimate $\bm{\varphi}$
\State \textbf{Sampling:}
\State Sample a Hamiltonian $\hat{H}\sim\GP$ over finite set~\cref{for:corH}
\State Compute approximation $\hat{H}^*$ of $\hat H$ and simulate GP-PHS
\end{algorithmic}
\end{algorithm}
\vspace{-0.2cm}
The complexity of the algorithm is dominated by the cost of training the GP and the sampling of the Hamiltonian $\hat{H}$, that is $\mathcal{O}(N^3+N^2n)$ with respect to the number $N$ of training points and the dimension $n$ of the state $\x$. The complexity can be reduced by inducing points methods, as in~\cite{wilson2015kernel}, and efficient sampling strategies as in~\cite{wilson2020efficiently}. As data-driven method, the accuracy of GP-PHS typically increases with the number of training points $N$. Thus, the choice of $N$ is a trade-off between computational complexity and accuracy of the prediciton. However, as shown in~\cref{cor:1}, the model output is physically plausible under all circumstances.
%%%%%%%%%%%%%%%%%%%%%%%%%%%%%%%%%%%%%%%%%%%%%%%%%%
%%%%%%%%%%%%%%%%%%%%%%%%%%%%%%%%%%%%%%%%%%%%%%%%%%
\section{Proof of concept}\label{sec:sim}
We consider the dynamics of an iron ball in the magnetic field of a controlled inductor
\begin{align}\label{for:sys}
        \dx&=\begin{bmatrix}
            0 & 1 & 0\\ -1 & -c\vert x_2\vert & 0\\0 & 0 & -1/R
        \end{bmatrix}\nabla_x H( \x)+\begin{bmatrix}
            0\\0\\1
        \end{bmatrix}u
\end{align}
with the vertical position of the ball $x_1$, its momentum~$x_2$ and the magnetic flux $x_3$. The electric resistance is assumed to be $R=0.1$ and the drag coefficient $c=1$. The Hamiltonian is given by $H(\x)=\frac{1}{2m}x_2^2+\frac{1}{2}\frac{x_3^2}{L(x_1)}$,
with the mass of the ball $m=0.1$, and the inductance $L(x_1)=1/(0.1+x_1^2)$. 
%\begin{figure}[b]
%\begin{center}
%	\input{figure/mag.tex}
%	\caption{Electromechanical system with an iron ball with mass $m$ and position $x$ in the magnetic flux $\Phi$ of a controlled inductor with resistance~$R$, inductance $L(x)$ and voltage $V(t)$.}\vspace{-0.2cm}
%	\label{fig:mag}
%\end{center}
%\end{figure}
Thus, the Hamiltonian establishes the coupling between the mechanical and electrical part of the system. For demonstration of learning the system~\cref{for:sys} with GP-PHS, we assume that the true Hamiltonian $H$ and the drag coefficient $c$ \emph{are unknown}. The training set $\D$ consists of data recorded from a single trajectory with initial condition $\x_0=[1,0,0]^\top$ and rectangular input signal $u$ as shown in~\cref{fig:mag_training}. From this trajectory, we collect every $0.05\si{\second}$ a time/state data pair which leads to a dataset of 401 training points with additive measurement noise $\mathcal{N}(0,0.01^2)$. We use a GP with squared exponential kernel to create a dataset where we minimize the NLML. To obtain the GP-PHS model with the unknown drag parameter $c=1$, we set $\Phi_J=\R_+$ and minimize the NLML, resulting in $\varphi_J=0.89$ as estimate for $c$. We sample 2142 equally distributed points for $\hat H$ of the joint distribution~\cref{for:corH} over the set $[-0.5,2]\times[-0.2,0.2]\times[-3,5]$ and use spline interpolation to achieve $\hat H^*$.

\Cref{fig:mss} shows the sampling and simulation of five trajectories of the GP-PHS model. Note that we use here 
\begin{figure}[bht]
\begin{center}
	\tikzsetnextfilename{mag_training}
\begin{tikzpicture}
\begin{axis}[
  name=plot1,
  ylabel={\sffamily State $\x$},
  legend pos=north west,
  width=\columnwidth,
  height=4cm,
    xticklabel=\empty,
 ymax=2.8,
  xmin=0,
  xmax=20,
      label style={font=\footnotesize},
  tick label style={font=\footnotesize},
  tick pos=left,
  legend style={font=\footnotesize},
  legend cell align={left},
  legend pos=north west,legend columns=3]
\addplot[color=green!60!black,line width=1pt,no marks] table [x index=0,y index=2]{data/meg_system_learning.dat};
\addplot[color=red,line width=1pt,no marks] table [x index=0,y index=3]{data/meg_system_learning.dat};
\addplot[color=blue,line width=1pt,no marks] table [x index=0,y index=4]{data/meg_system_learning.dat};
\legend{position $x_1$, momentum $x_2$, flux $x_3$};
\end{axis}
\begin{axis}[
name=plot2,
   at=(plot1.below south east), anchor=above north east,
  xlabel={\sffamily Time (s)},
  ylabel={\sffamily  Input $u$},
  legend pos=north west,
  width=\columnwidth,
  height=3cm,
  xmin=0,
  xmax=20,
      label style={font=\footnotesize},
  tick label style={font=\footnotesize},
  tick pos=left,
  legend style={font=\footnotesize},
  legend cell align={left}]
\addplot[color=black,line width=1pt,no marks] table [x index=0,y index=1]{data/meg_system_learning_u.dat};
\end{axis}
\end{tikzpicture} 
	\vspace{-0.4cm}\caption{Single training trajectory over time $t$ with input input $u$}\vspace{-0.7cm}
	\label{fig:mag_training}
\end{center}
\end{figure}
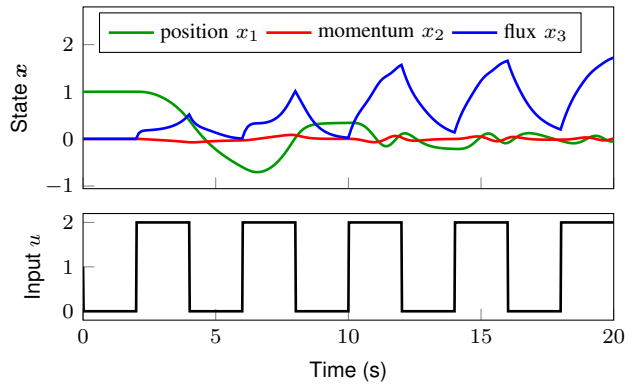
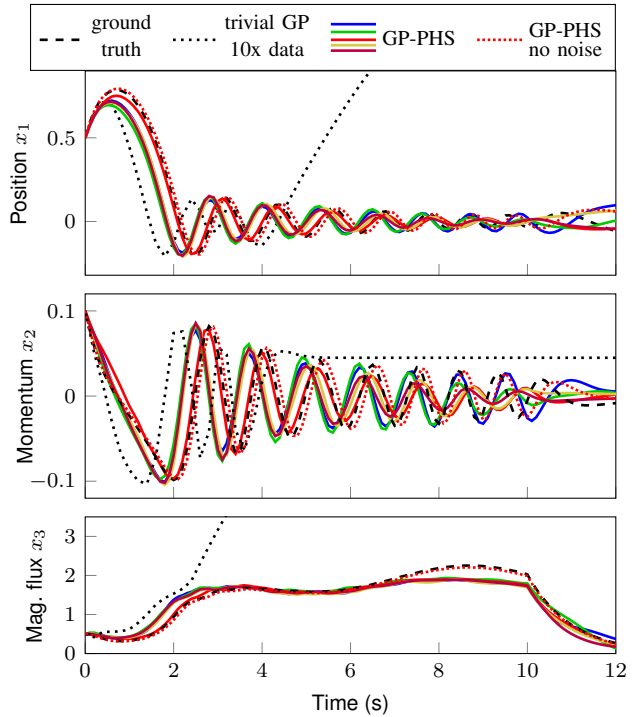
\begin{figure}[thb]
\begin{center}
	\tikzsetnextfilename{mag_gp}
\begin{tikzpicture}
\begin{axis}[
  name=plot1,
  ylabel={\sffamily Position $x_1$},
  legend pos=north east,
  width=\columnwidth,
  height=4.3cm,
  ymax=0.9,
  xmin=0,
  xmax=12,
  legend style={at={(1,1)},anchor=south east},
      label style={font=\footnotesize},
  tick label style={font=\footnotesize},
  tick pos=left,
  legend style={font=\footnotesize},
  xticklabels={,,},
legend style={row sep=-8pt},
  legend style={/tikz/every even column/.append style={column sep=0.2cm}},
  legend columns=5,transpose legend]
\addplot[color=white,line width=1pt,no marks] coordinates {(-100,0) (-100,0)};
\addplot[color=white,line width=1pt,no marks] coordinates {(-100,0) (-100,0)};
\addplot[color=black,dashed,line width=1pt,no marks] table [x index=0,y index=22]{data/meg_system_data.dat};
\addplot[color=white,line width=1pt,no marks] coordinates {(-100,0) (-100,0)};
\addplot[color=white,line width=1pt,no marks] coordinates {(-100,0) (-100,0)};
\addplot[color=white,line width=1pt,no marks] coordinates {(-100,0) (-100,0)};
\addplot[color=white,line width=1pt,no marks] coordinates {(-100,0) (-100,0)};
\addplot[color=black,dotted,line width=1pt,no marks] table [x index=0,y index=1]{data/meg_system_gp_more_data.dat};
\addplot[color=white,line width=1pt,no marks] coordinates {(-100,0) (-100,0)};
\addplot[color=white,line width=1pt,no marks] coordinates {(-100,0) (-100,0)};
\addplot[color=blue,line width=1pt,no marks] table [x index=0,y index=2]{data/meg_system_data.dat};
\addplot[color=green!80!black,line width=1pt,no marks] table [x index=0,y index=6]{data/meg_system_data.dat};
\addplot[color=red,line width=1pt,no marks] table [x index=0,y index=10]{data/meg_system_data.dat};
\addplot[color=yellow!80!black,line width=1pt,no marks] table [x index=0,y index=14]{data/meg_system_data.dat};
\addplot[color=purple,line width=1pt,no marks] table [x index=0,y index=18]{data/meg_system_data.dat};
\addplot[color=white,line width=1pt,no marks] coordinates {(-100,0) (-100,0)};
\addplot[color=white,line width=1pt,no marks] coordinates {(-100,0) (-100,0)};
\addplot[color=red,densely dotted,line width=1pt,no marks] table [x index=0,y index=1]{data/meg_system_gpphs_no_noise.dat};
\addplot[color=white,line width=1pt,no marks] coordinates {(-100,0) (-100,0)};
\addplot[color=white,line width=1pt,no marks] coordinates {(-100,0) (-100,0)};
\legend{ground,\phantom{a},\phantom{a},\phantom{a},truth,
trivial GP,\phantom{a},\phantom{a},\phantom{a},10x data, \phantom{a},\phantom{a},GP-PHS,\phantom{a},\phantom{a},\phantom{a},GP-PHS,\phantom{a},\phantom{a},no noise};
\end{axis}
\begin{axis}[
name=plot2,
   at=(plot1.below south east), anchor=above north east,
  ylabel={\sffamily Momentum $x_2$},
  ylabel shift = -9 pt,
  legend pos=north west,
  width=\columnwidth,
  height=4.3cm,
    xticklabel=\empty,
  ymin=-0.12,
  ymax=0.12,
  xmin=0,
  xmax=12,
      label style={font=\footnotesize},
  tick label style={font=\footnotesize},
  tick pos=left,
  legend style={font=\footnotesize},
  legend cell align={left}]
\addplot[color=blue,line width=1pt,no marks] table [x index=0,y index=3]{data/meg_system_data.dat};
\addplot[color=green!80!black,line width=1pt,no marks] table [x index=0,y index=7]{data/meg_system_data.dat};
\addplot[color=red,line width=1pt,no marks] table [x index=0,y index=11]{data/meg_system_data.dat};
\addplot[color=yellow!80!black,line width=1pt,no marks] table [x index=0,y index=15]{data/meg_system_data.dat};
\addplot[color=purple,line width=1pt,no marks] table [x index=0,y index=19]{data/meg_system_data.dat};
\addplot[color=black,dashed,line width=1pt,no marks] table [x index=0,y index=23]{data/meg_system_data.dat};
\addplot[color=black,dotted,line width=1pt,no marks] table [x index=0,y index=2]{data/meg_system_gp_more_data.dat};
\addplot[color=red,densely dotted,line width=1pt,no marks] table [x index=0,y index=2]{data/meg_system_gpphs_no_noise.dat};
%\legend{data-based optimization,closed-loop optimization};
\end{axis}
\begin{axis}[
name=plot3,
   at=(plot2.below south east), anchor=above north east,
  ylabel={\sffamily  Mag. flux $x_3$},
  xlabel={\sffamily  Time (s)},
  legend pos=north west,
  width=\columnwidth,
  height=3.4cm,
  ymin=0,
  ymax=3.5,
  xmin=0,
  xmax=12,
  tick pos=left,
      label style={font=\footnotesize},
  tick label style={font=\footnotesize},
  legend style={font=\footnotesize},
  legend cell align={left}]
\addplot[color=blue,line width=1pt,no marks] table [x index=0,y index=4]{data/meg_system_data.dat};
\addplot[color=green!80!black,line width=1pt,no marks] table [x index=0,y index=8]{data/meg_system_data.dat};
\addplot[color=red,line width=1pt,no marks] table [x index=0,y index=12]{data/meg_system_data.dat};
\addplot[color=yellow!80!black,line width=1pt,no marks] table [x index=0,y index=16]{data/meg_system_data.dat};
\addplot[color=purple,line width=1pt,no marks] table [x index=0,y index=20]{data/meg_system_data.dat};
\addplot[color=black,dashed,line width=1pt,no marks] table [x index=0,y index=24]{data/meg_system_data.dat};
\addplot[color=black,dotted,line width=1pt,no marks] table [x index=0,y index=3]{data/meg_system_gp_more_data.dat};
\addplot[color=red,densely dotted,line width=1pt,no marks] table [x index=0,y index=3]{data/meg_system_gpphs_no_noise.dat};
\end{axis}
\end{tikzpicture} 
	\vspace{-0.7cm}\caption{Five realizations of the GP-PHS and the resulting system trajectories. The realizations for the position $x_1$, momentum $x_2$ and the magnetic flux $x_3$ are significantly better approximations of the true system behavior (dashed line) than a trivial GP learning approach (black dotted).}\vspace{-0.2cm}
	\label{fig:mss}
\end{center}
\end{figure}
\hspace{-0.1cm}an initial condition $\x_0=[0.5,0.1,0.5]^\top$ and input function $u$ which was not seen before in the dataset $\D$. As baseline comparison, we trained a GP function with squared exponential kernel as a direct mapping from $\x\mapsto\dx$ without using the underlying PH structure. To achieve a considerable performance, we had to train this naive GP model on a dataset with ten times as many samples as for the GP-PHS, using multiple trajectories with random initial states. However, this baseline approach still does not generalize well as visualized in~\cref{fig:mss}, where the posterior mean of the trivial GP learning approach (dotted) cannot 
%\begin{figure}[h]
%\begin{center}
%\vspace{-0.3cm}
%	\input{figure/training_data.tex}
%	\caption{The training dataset consists of 401 data points in the state space generated by a single trajectory.}\vspace{-0.2cm}
%	\label{fig:training_data}
%\end{center}
%\end{figure}
reproduce the true system behavior. In contrast, the realizations of the proposed GP-PHS generalize well for the new initial condition and input function. For the sake of completeness, we include a sample (red dotted) of a GP-PHS where the measurements are not corrupted by any noise. 

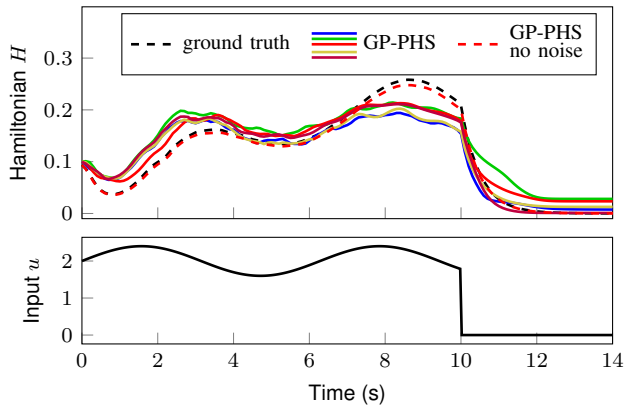
\begin{figure}[ht]
\begin{center}
\vspace{-0.2cm}
	\tikzsetnextfilename{ham_gp}
\begin{tikzpicture}
\begin{axis}[
name=plot3,
  ylabel={\sffamily Hamiltonian $H$},
  legend pos=north east,
  width=\columnwidth,
  height=4.4cm,
    xticklabel=\empty,
  ymin=-0.01,
  ymax=0.399,
  xmin=0,
  xmax=14,
    label style={font=\footnotesize},
  tick label style={font=\footnotesize},
  tick pos=left,
    legend style={font=\footnotesize},
legend style={row sep=-8pt},
  legend style={/tikz/every even column/.append style={column sep=0.2cm}},
  legend columns=5,transpose legend]
  \addplot[color=white,line width=1pt,no marks] coordinates {(-100,0) (-100,0)};
\addplot[color=white,line width=1pt,no marks] coordinates {(-100,0) (-100,0)};
\addplot[color=black,dashed,line width=1pt,no marks] table [x index=0,y index=25]{data/meg_system_data.dat};
\addplot[color=white,line width=1pt,no marks] coordinates {(-100,0) (-100,0)};
\addplot[color=white,line width=1pt,no marks] coordinates {(-100,0) (-100,0)};
\addplot[color=blue,line width=1pt,no marks] table [x index=0,y index=5]{data/meg_system_data.dat};
\addplot[color=green!80!black,line width=1pt,no marks] table [x index=0,y index=9]{data/meg_system_data.dat};
\addplot[color=red,line width=1pt,no marks] table [x index=0,y index=13]{data/meg_system_data.dat};
\addplot[color=yellow!80!black,line width=1pt,no marks] table [x index=0,y index=17]{data/meg_system_data.dat};
\addplot[color=purple,line width=1pt,no marks] table [x index=0,y index=21]{data/meg_system_data.dat};
\addplot[color=white,line width=1pt,no marks] coordinates {(-100,0) (-100,0)};
\addplot[color=white,line width=1pt,no marks] coordinates {(-100,0) (-100,0)};
\addplot[color=red,dashed,line width=1pt,no marks] table [x index=0,y index=4]{data/meg_system_gpphs_no_noise.dat};
\addplot[color=white,line width=1pt,no marks] coordinates {(-100,0) (-100,0)};
\addplot[color=white,line width=1pt,no marks] coordinates {(-100,0) (-100,0)};
%\legend{data-based optimization,closed-loop optimization};
\legend{\phantom{a},\phantom{a},ground truth,\phantom{a},\phantom{a},\phantom{a},\phantom{a},GP-PHS,\phantom{a},\phantom{a}, GP-PHS,\phantom{a},\phantom{a},no noise};
\end{axis}
\begin{axis}[
name=plot4,
   at=(plot3.below south east), anchor=above north east,
  xlabel={\sffamily  Time (s)},
  ylabel={\sffamily  Input $u$},
  legend pos=north west,
  width=\columnwidth,
  height=3cm,
  xmin=0,
  xmax=14,
  label style={font=\footnotesize},
  tick label style={font=\footnotesize},
  tick pos=left,
  legend style={font=\footnotesize},
  legend cell align={left}]
\addplot[color=black,line width=1pt,no marks] table [x index=0,y index=1]{data/meg_system_data_u.dat};
\end{axis}
\end{tikzpicture} 
	\vspace{-0.4cm}\caption{The Hamiltonians (the total energy in the system) are decreasing over time for a zero input ($t>10\si{\second})$ such that all realizations represent a physical correct system behavior.}\vspace{-0.4cm}
	\label{fig:ham_gp}
\end{center}
\end{figure}

Finally, the total energy of the GP-PHS samples are visualized in~\cref{fig:ham_gp}. All realizations of the GP-PHS represent physical plausible systems as the total energy is non-increasing for a zero input, i.e., for $t>10\si{\second}$.
%%%%%%%%%%%%%%%%%%%%%%%%%%%%%%%%%%%%%%%%%%%%%%%%%%
\section*{Conclusion}
In this paper, we introduce Gaussian process Port-Hamiltonian systems as a Bayesian learning approach for physical systems. The probabilistic nature of the model allows us to generate all possible realizations of a learned PHS under the GP prior based on a finite dataset. With the proposed PHS kernel, we prove that all realizations of the GP distribution respect the PHS structure. Finally, we show that GP-PHS share the interconnection property and passivity characteristic with PHS. A simulation highlights the superior behavior in contrast to trivial GP regression. In future work, we will develop energy based control strategies based on the proposed GP-PHS model.

%\appendices

%\section*{Acknowledgment}

\bibliography{mybib}
\bibliographystyle{ieeetr}

\end{document}